\documentclass[11pt]{article} 
\usepackage[preprint]{acl}

\usepackage{times}
\usepackage{latexsym}
\usepackage{booktabs}
\usepackage[linesnumbered,ruled,vlined]{algorithm2e}
\usepackage{amsmath}
\usepackage{amssymb}
\usepackage[skip=0pt]{caption}
\usepackage{pifont}
\usepackage{multirow} 
\usepackage{titling}
\usepackage{tabularx}
\usepackage[T1]{fontenc}

\usepackage[utf8]{inputenc}

\usepackage{microtype}

\usepackage{inconsolata}

\usepackage{graphicx}

\title{TRACE: Timely Retrieval and Alignment for Cybersecurity Knowledge Graph Construction and Expansion}

\author{
  Zijing Xu\textsuperscript{1}, 
  Ziwei Ning\textsuperscript{2}, 
  Tiancheng Hu\textsuperscript{3}, 
  Jianwei Zhuge\textsuperscript{1}, \\
  \textbf{Yangyang Wang}\textsuperscript{1}, 
  \textbf{Jiahao Cao}\textsuperscript{1}, 
  \textbf{Mingwei Xu}\textsuperscript{4}\thanks{~~Corresponding author.} \\
  \textsuperscript{1}Institute for Network Sciences and Cyberspace, Tsinghua University \\
  \textsuperscript{2}ZGC Laboratory \quad 
  \textsuperscript{3}SCS, Peking University \\
  \textsuperscript{4}Computer Science and Technology, Tsinghua University \\
}

\usepackage{bibentry}

\begin{document}

\maketitle

\begin{abstract}
The rapid evolution of cyber threats has highlighted significant gaps in security knowledge integration. Cybersecurity Knowledge Graphs (CKGs) relying on structured data inherently exhibit hysteresis, as the timely incorporation of rapidly evolving unstructured data remains limited, potentially leading to the omission of critical insights for risk analysis. To address these limitations, we introduce TRACE, a framework designed to integrate structured and unstructured cybersecurity data sources. TRACE integrates knowledge from 24 structured databases and 3 categories of unstructured data, including APT reports, papers, and repair notices. Leveraging Large Language Models (LLMs), TRACE facilitates efficient entity extraction and alignment, enabling continuous updates to the CKG. Evaluations demonstrate that TRACE achieves a 1.8$\times$ increase in node coverage compared to existing CKGs. TRACE attains the precision of 86.08\%, the recall of 76.92\%, and the F1 score of 81.24\% in entity extraction, surpassing the best-known LLM-based baselines by 7.8\%. Furthermore, our entity alignment methods effectively harmonize entities with existing knowledge structures, enhancing the integrity and utility of the CKG. With TRACE, threat hunters and attack analysts gain real-time, holistic insights into vulnerabilities, attack methods, and defense technologies.
\end{abstract}


\section{Introduction}
Knowledge graphs unify diverse data sources, providing semantic representations across formats and dimensions \cite{gopfert2025method}. 
They now serve as an external context for large language models (LLMs), improving domain-specific reasoning and relevance. In cybersecurity, such knowledge integration is essential for threat hunting, vulnerability tracking, and automated decision-making—capabilities increasingly supported by cybersecurity knowledge graphs (CKGs).
A CKG represents disparate threat intelligence repositories as a network of triples, enabling efficient path-based queries for detection and defense. STUCCO introduced an ontology for CKG construction \cite{iannacone2015developing}; UCO \cite{syed2016uco} extended this work by achieving deeper semantic consolidation across diverse datasets; BRON \cite{hemberg2024enhancements} developed a framework, which specifically integrates threat, vulnerability, and mitigation knowledge.

Automated techniques for extracting cybersecurity knowledge from unstructured text have become essential for the continuous evolution of CKGs~\cite{alharbi2025enhancing}. APT-specific extraction algorithms that combine deep learning with domain heuristics help build and maintain dynamic threat representations in CKGs~\cite{ren2022cskg4apt}. Meanwhile, structured behavior extraction methods have been applied to convert cyber threat intelligence (CTI) reports into structured representations of attack behaviors, enriching technical knowledge bases~\cite{li2022attackg}.

However, current CKGs exhibit several critical challenges, particularly given the growing complexity and heterogeneity of modern cybersecurity data sources:

(1) \textbf{\textit{Lacking high-coverage and generalizability.}} Existing CKG entity structures are incomplete. Many current approaches are built upon fixed ontologies that focus on a narrow set of cybersecurity entities and relations. Such designs hinder the ability to integrate new concepts and adapt to the growing complexity of cybersecurity knowledge. These rigid structures fail to align with and absorb emerging threat intelligence, resulting in limited scalability, weak generalization, and poor cross-domain interoperability.


(2) \textbf{\textit{Delayed updates hinder real-time applicability.}} The utility of a CKG depends on timely updates of entities and relations, as well as efficient entity alignment. However, many CKGs suffer from latency due to delays in source updates and the slow approval of vulnerability identifiers. Such delays are particularly critical for high-impact threats like zero-day vulnerabilities, where outdated or missing knowledge can result in significant security risks.

To address the challenges mentioned above, we propose a framework to \textbf{T}imely \textbf{R}etrieval and \textbf{A}lignment for \textbf{C}yber security knowledge graph's construction and \textbf{E}xpansion (\textbf{TRACE}). TRACE continuously updates and expands the CKG by retrieving entities and relations in diverse cybersecurity data sources, aligning them ontologically, and constructing new graph segments. TRACE integrates all data from 24 structured data sources and 3 unstructured data sources, including papers on attack and defense, advanced persistent threat (APT) reports, and repair notices. 

For the first challenge, we design a novel and extensible cybersecurity ontology to support the integration of heterogeneous data sources, particularly diverse types of unstructured information. Identical nodes from unstructured data sources are consolidated and merged into the current CKG by LLMs, and inter-database nodes are appropriately linked. 

For the second challenge, we deploy an automated update module for TRACE. This ensures that the latest cybersecurity intelligence and related knowledge are promptly integrated and semantically encoded in the CKG. To ensure semantic coherence across such diverse data sources, alignment maps extracted entities to existing nodes, enabling the timely and consistent integration of new knowledge. 

We evaluate the coverage of TRACE by comparing it with existing CKGs. Our findings indicate that TRACE contains $1.8\times$ more nodes than the largest previously known CKG. In addition, we assess TRACE's entity extraction performance, achieving a precision of 86.08\%, a recall of 76.92\%, and an F1 score of 81.24\%, representing a 7.8\% improvement over the best-known baseline utilizing LLMs. Furthermore, our evaluation of entity alignment demonstrates that our methods are highly effective in aligning extracted entities with those in existing knowledge graphs.

Our contributions are three-fold as follows:
\begin{itemize}
    \item We propose TRACE, a framework that constructs the largest cybersecurity knowledge graph to date, unifying 24 structured and 3 unstructured data sources, improving coverage by 1.8$\times$ over the previous state of the art.
\end{itemize}
\begin{itemize}
    \item We develop an automated approach to extract entities from unstructured data sources with LLMs. This method achieves a maximum F1 score of 81.24\%, representing a 7.8\% improvement over the baseline.
\end{itemize}
\begin{itemize}
    \item We define a generalized cybersecurity ontology and incorporate entity alignment to support continuous and adaptive expansion of CKG. By accommodating diverse entity types from structured and unstructured sources, the ontology enhances graph generalizability. 
\end{itemize}

\section{Related Work}
\subsection{Cybersecurity Ontology}
A cybersecurity ontology defines structured representations of entities and relations in the cybersecurity domain. 
It provides standardized definitions and semantic relationships among various elements such as assets, threats, vulnerabilities, attacks, and defensive measures. ~\cite{raskin2001ontology}. MITRE’s modular ontologies support rapid domain evolution through principles of modularity, clarity, and reusability~\cite{parmelee2010toward, obrst2012developing}.

The multilayered framework typically combines DOLCE-based upper ontologies with domain-specific ontologies to model foundational cybersecurity knowledge~\cite{oltramari2014building}. The Unified Cybersecurity Ontology (UCO) 1.0 integrated models from standards such as STIX 1.2, CVE, CCE, CVSS, CAPEC, and CYBOX~\cite{syed2016uco}, while UCO 2.0, based on STIX 2.0, further refined and extended this integration~\cite{pingle2019relext}. Both versions build upon the IDS ontology foundation~\cite{pinkston2003target}. In addition, specialized ontologies have been developed for specific tasks such as zero-day attack detection~\cite{razzaq2014ontology} and military cyber operation planning~\cite{chan2015ontological}.

In our work, TRACE integrates and extends these prior ontologies to establish a high-coverage and multidimensional cybersecurity ontology structure. This expansion broadens the scope of the CKG, enabling it to encompass a wider array of cybersecurity knowledge.

\subsection{Cybersecurity Knowledge Graph}
Recent CKG efforts integrate diverse data from structured and unstructured sources. Several methods have been proposed that leverage heterogeneous inputs such as CVE, CWE, and CPE, with a focus on textual descriptions and associated attributes~\cite{jia2018practical, qin2019automatic, sun2020method}. SecTKG \cite{sun2023sectkg} is an automated framework that builds knowledge graphs from open-source security tools; AISecKG \cite{agrawal2023aiseckg} constructed a knowledge graph for cybersecurity education.

In CKG applications, OpenCTI \cite{ruohonen2024instrumenting} serves as an open-source platform aiding organizations in CTI data management based on CKG. 
CTINexus \cite{cheng2024ctinexus} automatically constructs a CTI-centric CKG using LLMs.
MalONT2.0 \cite{christian2021ontology} constructed a CKG from Android malware reports using semantic and syntactic features. Knowledge graph embedding models have been explored for entity completion in vulnerability graphs \cite{host2022constructing}, while path-based methods have been applied to infer new relationships via entity connectivity analysis \cite{jia2018practical}.

Existing CKGs often exhibit limited coverage and lack timeliness due to the latency in structured data sources. TRACE addresses these limitations by aggregating a broad spectrum of data sources and promptly integrating entities extracted from the latest unstructured texts, thereby enhancing the comprehensiveness and timeliness of the CKG.

\begin{figure*}[t]
  \centering
  \includegraphics[width=0.88\textwidth]{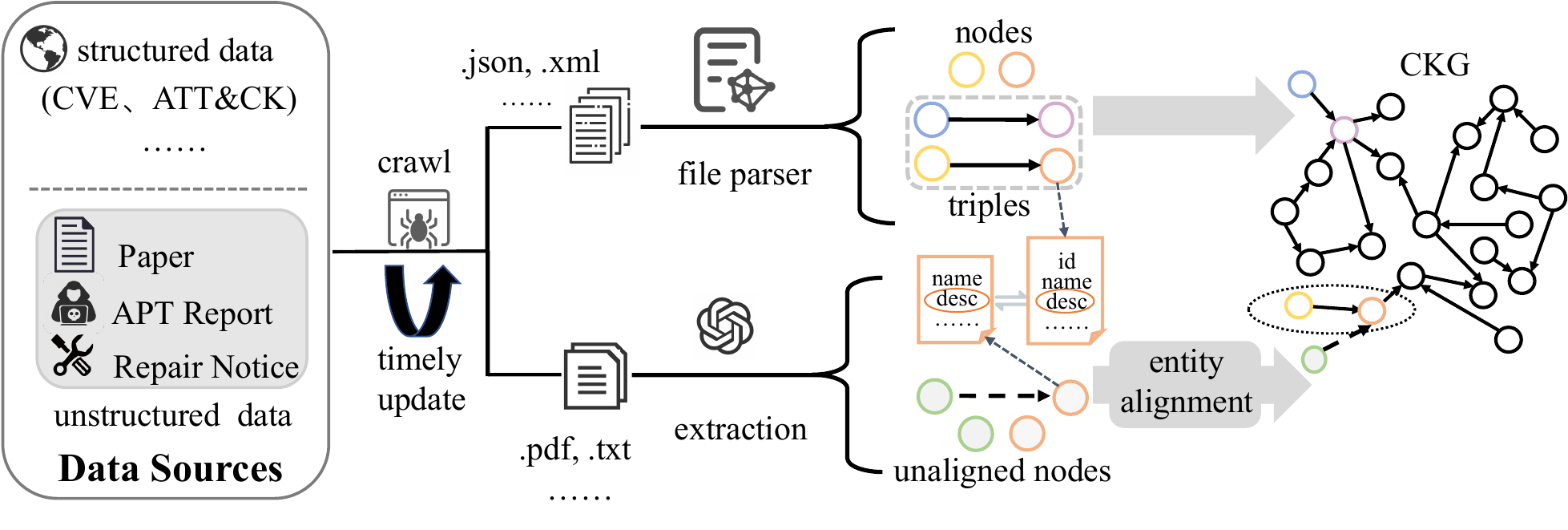}
  \caption{Framework of TRACE}
  \label{overview}
\end{figure*}

\section{System Architecture}
The framework of TRACE consists of three key modules: Information Acquisition, filtering and validation, entity standardization and alignment, as depicted in Figure \ref{overview}.



\subsection{Construction of the CKG}
In TRACE, the ontology is structured into three primary dimensions: vulnerability, attack, and defense. Entities extracted from diverse data sources are categorized under these dimensions, whose properties are mapped to corresponding subsets of STIX Domain Objects (SDOs). This alignment facilitates a comprehensive representation of specialized cybersecurity knowledge domains. 


\subsubsection{Entity Types}
To ensure the standardization of the graph, entity types extracted from structured and unstructured data sources are processed independently. These nodes maintain the knowledge types as well as the hierarchical structure inherent in the data sources. When the subcategory count is minimal, the hierarchical structure is flattened into a single level, with subtitled labels retained to differentiate subcategories. Conversely, a distinct node type is created when the subcategory count is substantial. These new node types inherit the name of the parent structure, and their contents comprise elements drawn from the original sublevel structure.

Unstructured data comprises textual documents that may include diagrams or code snippets. We categorize unstructured data sources into three primary types: APT reports, repair notices, and papers on attack or defense. Based on the focus of each type, we adopt the MITRE ATT\&CK \cite{strom2018mitre} framework and tailor our ontology schemas accordingly. For instance, an APT report typically describes how a specific advanced persistent threat leverages targeted techniques to exploit one or more vulnerabilities against a given system.
So we extract entity types involving "vulnerabilities", "tools", "techniques", "groups", and "assets".

\subsubsection{Relation Types}
Similarly, the establishment of relations in structured and unstructured data sources needs to be handled separately. All pre-existing relations in structured data sources are preserved. Additionally, newly created sub-level nodes are linked to their corresponding original data nodes to represent a "belongs-to" relationship.

Relation types for unstructured data are defined based on the semantic logic of the source text. Academic papers typically document the utilization of specific methods or tools in the development of novel attack techniques, which are deployed against targeted assets to uncover known or previously unreported vulnerabilities. Finally, they recommend suitable mitigation strategies, offering guidance for defenders responsible for remediation. Consequently, the defined relations involve "discovers", "uses", "causes", "reflects", "mitigates", and "solves".

\subsection{Collection and Processing}
\begin{algorithm}
\caption{Collection and Processing}\label{algorithm}

\KwData{ $U$ is a list of unstructured texts $\{u_1, u_2, … , u_n\}$; $KG$: knowledge graph; $N$, $TP$: nodes, triples in $KG$; $\theta$, $\delta$ for similarity, time interval threshold; $time_c$, $time_l$ for current, last timestamp}
\KwResult{Updated $N$ and triples $TP$}
$N, TP \gets \emptyset$; $S \gets \emptyset$; $i, j, m \gets 0$

\While{$time_c - time_l = \delta$}{
    $/$$/$Structured Acquisition\newline
    \While{$E_i.timestamp \leq time_c$}{
        $N_s = GetNode_{\text{mongoDB}}(E_i)$\\
        $TP_s = GetTriples_{\text{mongoDB}}(E_i)$\\
        $i = i + 1$\;
    }
    $/$$/$Unstructured Acquisition\newline
    \ForEach{$u_j \in U$}{
        \If{$u_j \notin \text{paper}$}{
        $S = S \cup \{u_j\}$\;
        }
        \Else{
        \If{$u_j.\text{security} = \text{true}$}{
            $S = S \cup \{u_j\}$\;
        }
    }
    }
    \ForEach{$s_k \in S$}{
        $N_u = GetNode_{\text{LLM}}(s_k)$\;
        $TP_u = GetTriples_{\text{LLM}}(s_k)$\;
    }
    $/$$/$Filtering and Validation\newline
    $N_u = \{n \in N_u \mid n \neq \emptyset\}$\;
    $/$$/$Entity Standardization and Alignment\newline
    $EntityAlignment(N_u, TP_u)$\;
    $N = N \cup N_s \cup N_u$\;
    $TP = TP \cup TP_s \cup TP_u$\;
    $KG = KG \cup getkg(N, TP)$\;
    $time_l = time_c$\;
}
\end{algorithm}
\subsubsection{Structured Information Acquisition}
The raw content in structured data sources must be parsed and decomposed before entity extraction, as shown in lines 3–7 of Algorithm~\ref{algorithm}.  

For structured data sources, we implement an automated pipeline to retrieve, update, and organize data from each repository. We begin by fetching the initial files from each source website or database. We truncate fields and apply tailored regular expressions to extract the desired properties in each entity’s JSON document. All property names are normalized to a unified schema. It is worth mentioning that TRACE records each source’s timestamp and crawls only data newer than the previous run, which is the basis for automated updates.

TRACE employs a plugin-based architecture to manage each data source independently. For each source, TRACE uses \textbf{Full crawls} and \textbf{Incremental crawls} to acquire data concurrently. \textbf{Full crawls}, which are executed at longer intervals, systematically retrieve all available content to avoid missing releases that exceed typical pagination limits; \textbf{Incremental crawls}, which are executed more frequently, poll recent pages for updates to ensure the CKG reflects the latest data in near real time. After collection, TRACE applies rigorous validation and deduplication procedures to all entities and triples, guaranteeing both completeness and consistency before integrating them into the graph.  

\subsubsection{Unstructured Information Acquisition}
For unstructured data sources, the data preprocessing pipeline begins with format conversion and cleansing, followed by a preliminary screening process to validate the presence of required content within each document. The filtered texts are sent to the LLM for detailed knowledge extraction, as shown in lines 8-17 of Algorithm~\ref{algorithm}. 

APT reports and repair notices are sourced from annual disclosures and periodic product updates, ensuring that essential entities and triples are available. Since only a subset of offensive and defensive papers contains extractable cybersecurity insights, we first apply an LLM–based binary classifier on each paper’s title and abstract to distinguish relevant from irrelevant texts. Only those papers deemed relevant proceed to the subsequent knowledge extraction stage.

The information acquisition has a two‑step strategy: first, \textbf{extract} candidate nodes from each document; second, \textbf{combine} the nodes based on the predefined relationship schema. The combination is then submitted back to the LLM to validate the existence of each relation in the source text. 

We leverage few‐shot learning with Retrieval‐Augmented Generation (RAG) and target regularization to maximize node extraction accuracy. By our prompt‐engineering framework, each node category and triple pattern is defined systematically and illustrated with examples. For nodes with standardized formats, such as CVE identifiers, we apply regex matching for direct extraction. For attributes exhibiting contextual semantic variability, we maintain a curated retrieval repository of previously extracted nodes and triples, which serves as a reference database for LLMs.

Before generating entity extraction results through prompts that provide formal definitions and task-specific exemplars, LLM uses the repository to retrieve relevant information, thereby enhancing the model's comprehension of contextual requirements. This structured guidance enables LLM to generate extraction results in standardized formats while maintaining lightweight computational requirements with high efficiency.

\subsubsection{Filtering and Validation}
Characters that cause graph anomalies need to be replaced or deleted, such as "\textbackslash\textbackslash" and "--". The generated nodes without any other associations are considered \textbf{\textit{Isolated Nodes}}. The names of some \textbf{\textit{Isolated Nodes}} only contain serial numbers without valid information. These nodes will be filtered to reduce incoherence in the knowledge graph, as shown in line 22 of Algorithm~\ref{algorithm}.

\subsubsection{Entity Standardization and Alignment}
To ensure the correct incremental update of the graph, all data must be standardized into a consistent format. This involves organizing the content of each node, ensuring uniformity across all nodes while also preserving the uniqueness of nodes from different data sources. To reduce hallucinations in subsequent LLMs, it is crucial to maintain consistent node formatting, including ID prefixes, property names, and other standardized elements.
All nodes must include several common parameters to ensure consistency across the graph. Taking vulnerability dimension for instance, one data source may use a parameter "desc", while another may use "description". These differences need to be standardized. Additionally, every node must contain a "source" parameter, preventing confusion caused by nodes with the same name or type across different data sources. 
To support update tracking and deduplication, all nodes must include a timestamp parameter to record the exact collection time. This timestamp is continuously refreshed as TRACE executes its tasks, ensuring the graph always reflects the most recent collection. 

As shown in lines 23-27 of Algorithm~\ref{algorithm}, to prevent information redundancy between newly extracted entities and existing ones in the graph, these extracted entities from unstructured data sources must undergo entity alignment with their corresponding counterparts of identical types. We encode textual descriptions of the target node and its graph-based counterparts into vector embeddings, then select the top 20 candidate entities using similarity metrics.
If the highest similarity is lower than the value of $\theta$ we set, we consider it to be a new node and update it to the graph; if there exists at least one similarity higher than $\theta$, put them into the LLM for judgment to obtain the node with the highest similarity. Each node extracted needs to be aligned with the same type of node in the original graph. All edges originating from or pointing to the target node must be migrated to its highest-similarity counterpart through edge redirection.

\section{Implementation}
To construct the CKG, TRACE integrates data from a total of 27 sources, comprising 24 structured and 3 unstructured datasets. The structured data sources encompass four primary categories: vulnerability databases, vulnerability exploitation platforms, attack technique frameworks, and cybersecurity community platforms. For unstructured data sources, we selected 10,205 papers, 869 APT reports, and 6,784 repair notices published in the last decade. 

We systematically aggregated multi-dimensional corpora through diverse approaches according to the ontology structure we proposed, subsequently organizing them into structured nodes and triples. We employed MongoDB as the storage database, utilizing multiple built-in data types to accommodate various data formats. The nodes and triples extracted were subsequently integrated into the graph through alignment and merging to enhance its academic relevance.

\section{Experiments}
In this section, we focus on answering the following three research questions (RQs):
\begin{itemize}
    \item \textbf{RQ1}: How comprehensive is the coverage of the CKG constructed by TRACE, and to what extent are its nodes interconnected?
    \item \textbf{RQ2}: How effective are TRACE’s entity extraction and alignment regarding accuracy and completeness?
    \item \textbf{RQ3}: How do cybersecurity researchers and analysts leverage TRACE in practical case studies to validate its utility in real-world threat investigation?
\end{itemize}

\subsection{RQ1: Coverage and Connectedness }

\subsubsection{Coverage of CKG}

As for now, TRACE has successfully established 4,741,428 nodes and 24,980,064 edges in the current knowledge graph, which consists of 56 node types and 112 edge types. 

We compare the number of nodes and edges in the CKG with those in the largest known structured and unstructured cybersecurity knowledge graphs, BRON \cite{hemberg2024enhancements} and CSKG4APT \cite{ren2022cskg4apt}, as shown in Table~\ref{tab:coverage}. Our analysis revealed that ours contains $1.82\times$ more nodes and $1.79\times$ more edges than the previously largest CKG. Additionally, its node types and edge types represent $4.67\times$ and $11.2\times$ increases, respectively. These findings underscore the superior coverage of TRACE in representing cybersecurity knowledge.

\begin{table}[b]
\centering
\begin{tabular}{ccc}
\hline
                 & nodes              & edges               \\ \hline
BRON             & 500,000            & 5,060,000           \\
CSKG4APT         & 2,608,327          & 13,935,201          \\
\textbf{Our CKG} & \textbf{4,741,428} & \textbf{24,980,064} \\ \hline
\end{tabular}
\caption{The number of nodes and edges in CKGs.}
  \label{tab:coverage}
\end{table}

\subsubsection{Node Association Density Analysis}
\begin{table}[]
\centering
\begin{tabular}{ccc}
\hline
\textbf{all} & \textbf{Isolated Node} & \textbf{One-Edge Node} \\ \hline
4,741,428    & 124,866                 & 2,197,241               \\ \hline
\end{tabular}
\caption{The number of different kinds of nodes.}
  \label{tab:nodes}
\end{table}

Table~\ref{tab:nodes} illustrates the nodes' connection density.
\textbf{\textit{Isolated Nodes}} (nodes without any edges) accounting for 2.63\% of all nodes in the CKG. These nodes remain disconnected due to their failure to align with existing entities in the graph. Child nodes frequently evolve into \textbf{\textit{One-Edge Nodes}} (nodes with only one edge, comprising 46.34\% of total nodes) that exclusively associate with their parent entities. As elaborated in \textbf{Relation Types}, a substantial proportion of child nodes directly contributes to generating most of the \textbf{\textit{One-Edge Nodes}}.

In contrast, \textbf{\textit{Super Nodes}} represent highly interconnected entities with dense edge connections. We investigated the \textbf{\textit{Super Nodes}} with the most edges in the entire graph. For instance, CWE-79 has 32,396 edges associated with it, making it the node with the most edges in the entire graph. CWE-79 refers to Improper Neutralization of Input During Web Page Generation, or Cross-site Scripting. As the first of the 2024 CWE Top 25, the wide relevance of CWE-79 directly reflects its core position in the real threat ecosystem, and to a certain extent, reflects the connection between the vulnerability dimension and itself and other dimensions.

\begin{figure}
  \centering
  \includegraphics[width=0.47\textwidth]{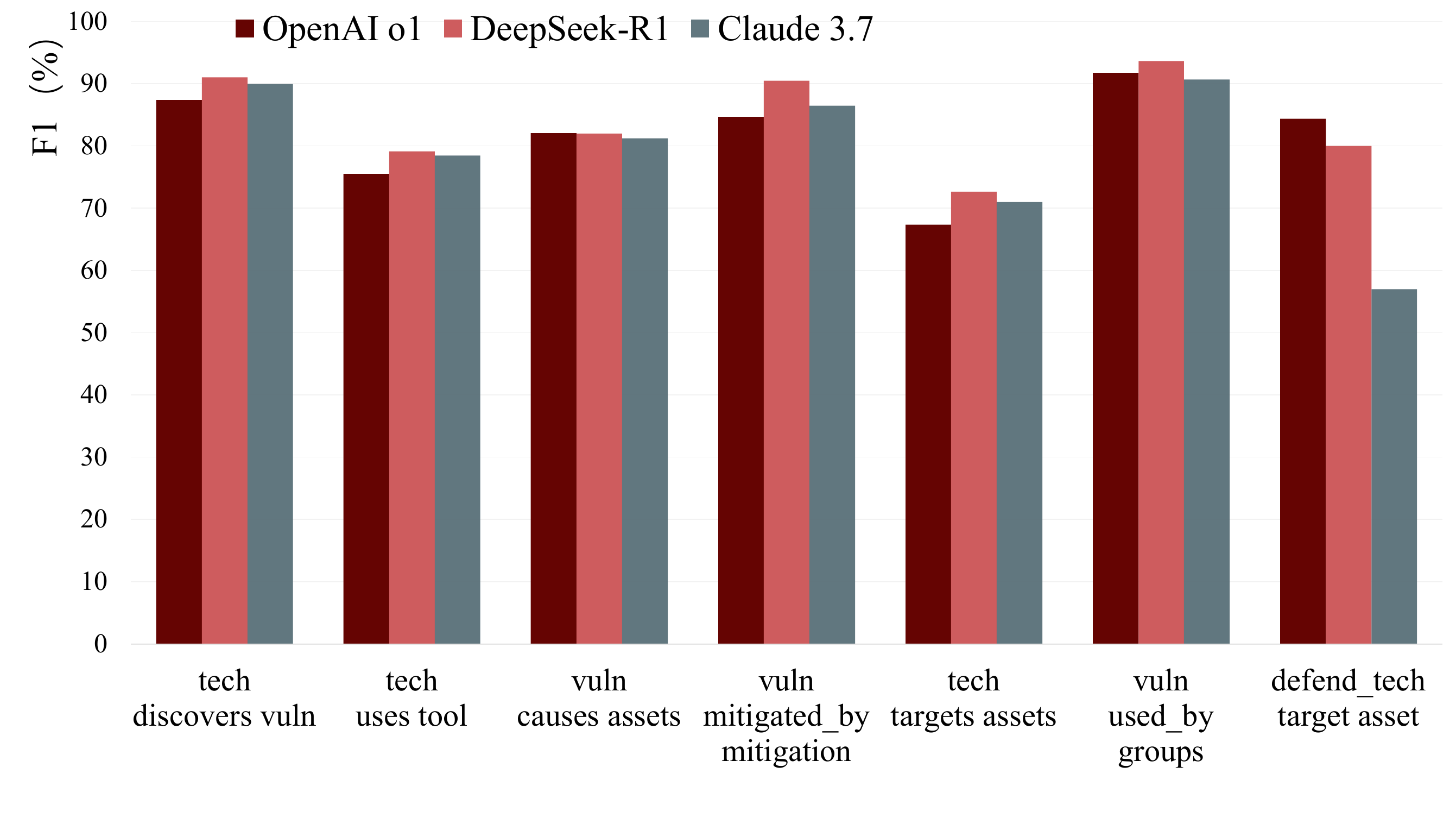}
  \caption{F1-score comparison of triple extraction in unstructured data sources across LLMs.}
  \label{graph:triples}
\end{figure}

\begin{table*}[]
\centering
\begin{tabular}{cccccc}
\hline
\textbf{Relationship}                    & \textbf{\#Src} & \textbf{\#Dst} & \textbf{\#Edges} & \textbf{Edges} & \textbf{E/P} \\ \hline
analytics \textit{consist} implementations   & 109            & 271            & 29,539            & 19,271        & 6.52E-01     \\
vuln \textit{affect} version                 & 1,569,668        & 1,127,590        & 1.77E+12         & 6,929,355      & 5.61E-07     \\
vuln \textit{has\_cvss} score                & 1,569,668        & 1,012,354        & 1.59E+12         & 992,279       & 6.24E-07     \\
vuln \textit{belong\_to} cwe                 & 1,569,668        & 1,430           & 2.24E+09         & 220,013       & 9.80E-05     \\
cpe \textit{belong\_to} infras               & 118,061         & 5              & 590,305           & 14,943        & 2.53E-02     \\
sensor \textit{map} data\_model              & 7              & 862            & 6,034             & 5,758         & 9.54E-01     \\
vuln \textit{exploited\_by} exp              & 1,569,668        & 236,572         & 3.71E+12         & 32,784        & 8.83E-08     \\
defend\_technique \textit{counter} technique & 182            & 1,043           & 189,826           & 646          & 3.40E-03     \\
cwe \textit{used\_by} attack\_pattern        & 1,430           & 615            & 879,450           & 737,649       & 8.39E-01     \\
group \textit{uses} technique                & 181            & 1,043           & 188,783           & 154,267       & 8.17E-01     \\ \hline
\end{tabular}
\caption{Triples Density, Relationship is the name of the triples collections, \#Src is the number of nodes in the type of start entity, \#Dst is the number of nodes in the type of end entity. \#Edges is the number of edges that should exist, Edges is the unique number of edges, EP is the ratio of Edges and \#Edges.}
  \label{tab:triples}
\end{table*}

\subsubsection{Triple Association Density Analysis}
Table~\ref{tab:triples} presents the qualitative statistical data of nodes and triples in the current graph. The observed-to-expected ratios of edges associated with the "vuln" entity are significantly lower than anticipated. This discrepancy arises because entities linked to "vuln" exhibit uniqueness in their structural definitions. For instance:

vuln \textit{affect} version: The "version" entities are decomposed from parent entities and thus predominantly correlate only with their parent entities.

vuln \textit{has\_cvss} score: "Scores" derived from CVSS vary substantially across vulnerabilities due to differences in attack vectors, impact scope, and exploit conditions.

The “E/P” values of most triples exceed 0.8, indicating that the CKG has strong connectivity.

\subsection{RQ2: Entity Extraction And Alignment}
\subsubsection{Entity Extraction }
\begin{table*}[ht]
  \centering
  \begin{tabularx}{0.97\textwidth}{c *{6}{>{\centering\arraybackslash}X}}
    \toprule
                & Ma$_{paper}$ & Mi$_{paper}$ & Ma$_{APT report}$ & Mi$_{APT report}$ & Ma$_{repair notice}$ & Mi$_{repair notice}$ \\
    \midrule
    GPT-NER     & 68.36\%      & 68.12\%      & 69.51\%           & 68.45\%           & 69.60\%              & 68.18\%               \\
    Cp-NER      & 62.54\%      & 62.48\%      & 63.72\%           & 63.96\%           & 64.44\%              & 63.89\%               \\
    ProML       & 66.18\%      & 67.89\%      & 70.34\%           & 69.02\%           & 73.44\%              & 68.51\%               \\
    CyberNER-FT & 67.64\%      & 66.72\%      & 67.45\%           & 67.11\%           & 70.84\%              & 66.39\%               \\
    \textbf{Our Method} &
      \textbf{73.57\%} & \textbf{73.25\%} & \textbf{73.01\%} & \textbf{76.80\%} &
      \textbf{81.24\%} & \textbf{80.87\%}      \\
    \bottomrule
  \end{tabularx}
  \caption{Comparison of entity extraction in unstructured data sources, Ma$_{type}$ and Mi$_{type}$ is the $F1_{\operatorname{macro}}$ and $F1_{\operatorname{micro}}$ of the current type of unstructured data sources respectively.}
  \label{tab:comparison}
\end{table*}

We randomly selected 200 texts each from APT reports, papers, and repair notices to evaluate entity extraction. Performance was assessed using macro- and micro-average F1-scores. The macro-average F1 is the mean of F1-scores computed separately for each entity type, while the micro-average F1 is the mean obtained by considering all entity instances across different text genres. Equations \ref{con:marco} and \ref{con:micro} present the respective calculation methods.

\begin{equation}
    F1_{\mathrm{macro}} = \frac{1}{n} \sum_{i=1}^{n} \left(
\frac{2 \cdot TP_i}{2 \cdot TP_i + FP_i + FN_i}
\right)\label{con:marco}
\end{equation}

\begin{equation}
    F1_{\operatorname{micro}}=\frac{2\sum_{i=1}^{n}TP_i}{2\sum_{i=1}^{n}TP_i+\sum_{i=1}^{n}(FP_i+FN_i)}\label{con:micro}
\end{equation}

We selected four state-of-the-art NER models: GPT-NER\cite{wang2023gpt}, Cp-NER\cite{chen2023one}, ProML\cite{chen2022prompt}, CyberNER-FT\cite{srivastava2023study}. We compared the experimental results of the various methods on both the macro‐average and micro‐average, as shown in Table~\ref{tab:comparison}. Because APT reports and repair notices tend to exhibit greater structural regularity and often contain publicly enumerated identifiers that are easier to extract, all five methods achieve higher scores on these genres than on papers. By combining few‐shot learning with an RAG framework, our method leverages both small‐sample learning and retrieval‐based techniques to achieve superior entity extraction performance.

We randomly select 20 texts from each of the three text types, which means 60 texts in total, and arrange the entities extracted into combinations according to the predefined triple types. We then apply holdout cross-validation to implement binary judgment of triples. The dataset is split into training, validation, and test sets in an 8:1:1 ratio, with the three LLMs that achieved the best binary classification performance in the paper used for triple judgment. 
To ensure fair comparison across models and mitigate the influence of stochastic variation, we conducted each experiment three times using fixed random seeds across all models. The final extraction for each model was obtained by taking the union across the three runs. The results are shown in the Figure \ref{graph:triples}. Finally, we use the best-performing model to classify the remaining relations, obtaining the list of entities and triples extracted from the unstructured data sources.

\subsubsection{Entity Alignment}
We regularly update and maintain the graph’s global list of entities and triples. For each new entity extracted from unstructured data, we locate all entities of the same type in the list and compute the semantic similarity of their "description" attributes using sentence-transformers, which can capture contextual information and long-distance dependencies, and thus offer more accurate semantic-similarity measures. After fine-tuning, we set the similarity threshold $\theta$ to 0.9. Only nodes whose similarity exceeds $\theta$ are passed to the next stage. We then apply zero-shot prompting to identify the final matching node, replacing the original node's properties in the extracted list with those of the matched node. Finally, we merge the revised entities and triples back into the existing global list.

\begin{figure}
  \centering
  \includegraphics[width=0.5\textwidth]{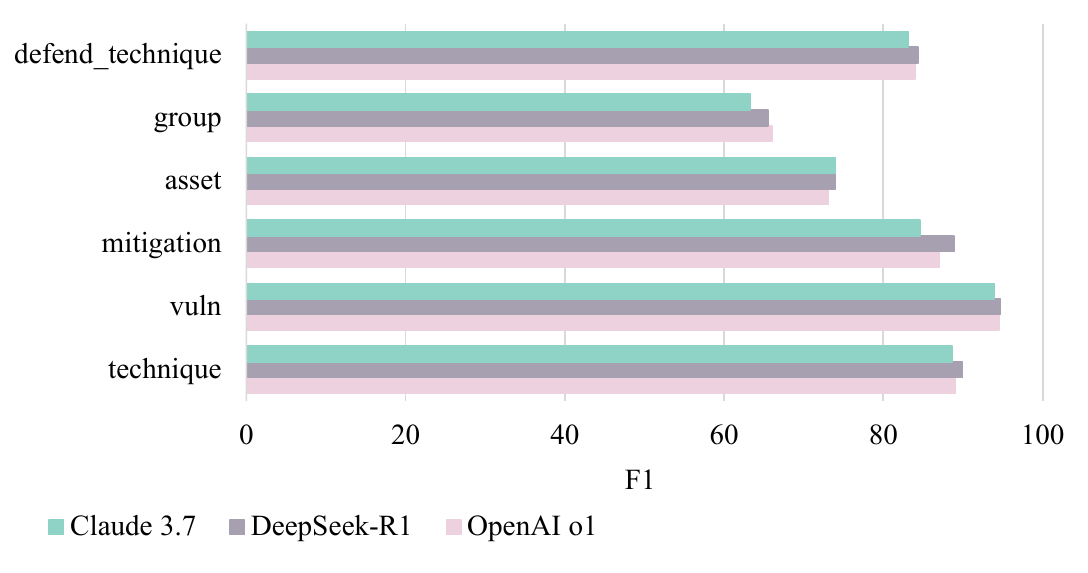}
  \caption{F1-score comparison of entity alignment in unstructured data sources across LLMs.}
  \label{graph:alignment}
\end{figure}

We randomly choose 50 instances for each of the six entity types and ask our expert team to judge how many of those matchings are correct, the results are shown in Figure \ref{graph:alignment}. We find that the F1-score for vuln entities is the highest because vuln identifiers follow well-defined formats that can be captured by regular expressions. If two descriptions share the same ID, they match directly without semantic textual similarity. By contrast, group entities have the lowest F1-score. We find that many groups operate covertly and employ a wide variety of attack strategies, so their descriptions are highly diverse, making entity alignment for group entities especially difficult.

\begin{figure*}
  \centering  \includegraphics[width=0.88\textwidth]{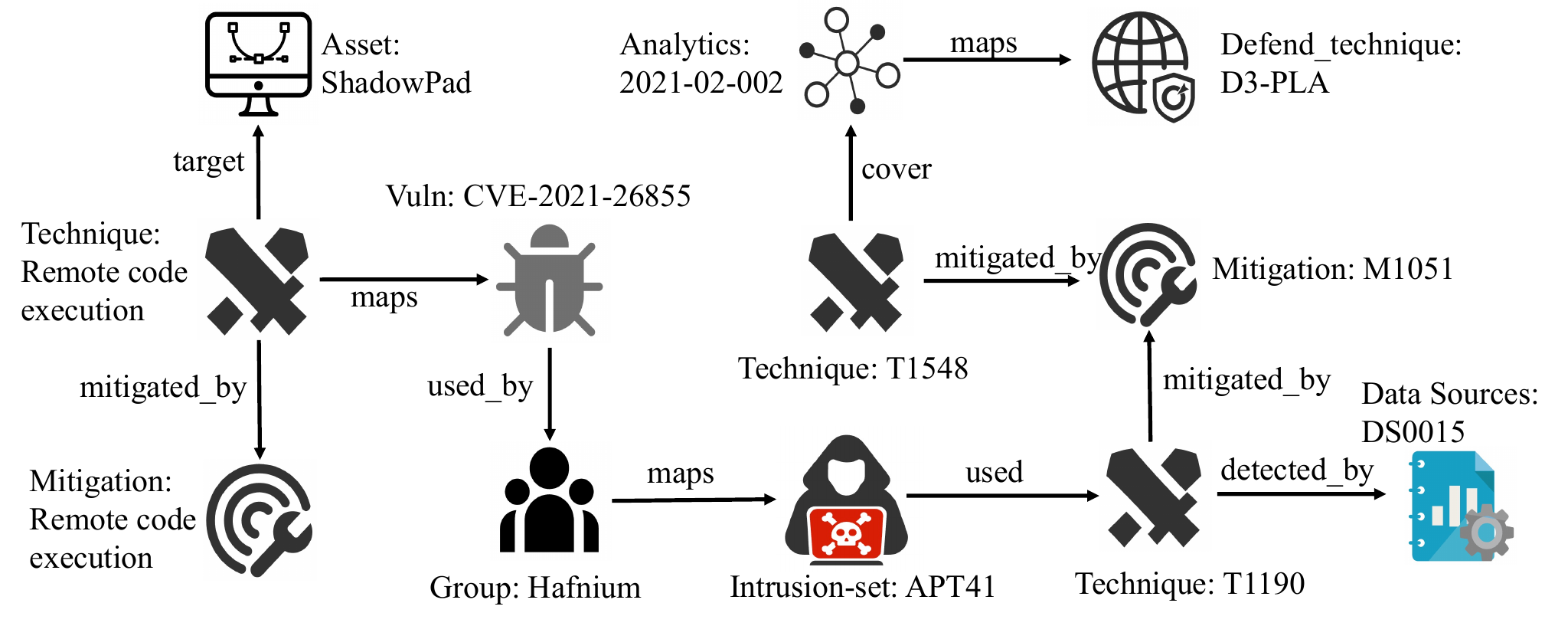}
  \caption{A sample of attack analysis using TRACE}
  \label{graph:casestudy}
\end{figure*}

\subsection{RQ3: Case Study}
We employ TRACE as the role of the attack analysts to illustrate how it bridges structured and unstructured data sources. 

\subsubsection{Connectivity between Entities}
As shown in Figure~\ref{graph:casestudy}, TRACE identifies the ProxyLogon vulnerability in Microsoft Exchange (CVE-2021-26855), which permits attackers to bypass authentication and assume administrator privileges \cite{khang2021botnet}. Analysis of APT reports attributes this exploit to the Hafnium group \cite{lechtik2021ghostemperor}, known in ATT\&CK as APT41. Subsequently, other threat actors such as ToddyCat (TG-3390) have also targeted this vulnerability. Within the ATT\&CK framework, APT41’s attack sequence includes technique T1190 (Exploit Public-Facing Application), detected by data source DS0015 (events collected by third-party services such as mail servers or web applications), and technique T1548 (Abuse Elevation Control Mechanism), for which M1051 is recommended as mitigation. In our defensive knowledge bases, these map respectively to CAR-2021-02-002 (Get System Elevation) and D3-PLA (Process Lineage Analysis) successively. When tracing an actual intrusion path, a single node in the graph may spawn multiple branches, allowing analysts to navigate and drill down to the precise threat intelligence required.

\subsubsection{The importance of Unstructured Data}
During our experiments, we observed that numerous entities extracted from unstructured data sources could not be matched to existing nodes in the CKG. This discrepancy arises because such information has not yet been incorporated into structured data sources. For instance, the OrpaCrab backdoor, detailed in a November 2024 APT report \cite{orpacrab}, employs the MQTT protocol for command and control, utilizes AES encryption to obfuscate its configuration, and leverages DNS over HTTPS to resolve its C2 domain, thereby evading traditional DNS monitoring. This malware was identified in a Gasboy fuel management system and was named "OrpaCrab" within that specific report. Consequently, in the CKG, it exists as a \textbf{\textit{One-Edge Node}} associated with the group, lacking broader integration. Similarly, several vulnerabilities have been disclosed in recent papers, with authors indicating that they have reported these issues to relevant organizations. However, these vulnerabilities have yet to receive official identifiers. Such entities are critical for risk analysis, as attackers may exploit these vulnerabilities before they are patched. Therefore, extracting entities from unstructured data sources and aligning them within the CKG is essential for timely and comprehensive threat assessment.

\section{Conclusion}
We introduce TRACE, a framework with a multidimensional cybersecurity ontology structure to enhance the construction and expansion of the CKG. TRACE integrates 24 structured data sources and 3 categories of unstructured data sources, constructing a CKG that encompasses 56 distinct node types and 112 edge types. 
By leveraging LLMs for efficient entity extraction and alignment, TRACE enhances the CKG's completeness and timeliness, providing a unified and dynamic view of cybersecurity information for security analysts and researchers with the latest knowledge. In the future, we plan to continuously expand the ontology structure, aiming to develop a more unified model of cybersecurity knowledge.

\section*{Limitations}
There are several limitations that we would like to address in future work.

\begin{itemize}
  \item TRACE integrates data from 24 structured data sources and 3 types of unstructured data sources, yielding a CKG with over four million nodes. Nevertheless, a non-negligible proportion of isolated nodes remains, revealing unlinked entities within the cybersecurity domain. We will continually expand TRACE’s data sources to drive further graph growth and improve its connectivity and application effectiveness.
  
  \item We leverage LLMs augmented with RAG for entity extraction from unstructured sources. We find that hallucinations in LLM outputs remain unavoidable, leading to errors such as misclassifying “tool” entities as “technique”. These misclassifications are due to ambiguous class boundaries and unbalanced training data distributions. Future work will refine prompt definitions, broaden few-shot examples, and rebalance training samples to mitigate these effects.
  
  \item Although we have implemented entity extraction, images in texts often contain critical information (e.g., exploit diagrams, technical schematics) that remains unexploited. We plan to adopt multimodal vision–language pretraining to tightly align visual and textual representations, enabling accurate interpretation of non-textual content in future work.
\end{itemize}

\bibliography{aaai2026}
\end{document}